# Photon number conservation in time dependent systems

**J.B. PENDRY**[*1]

[1]*The Blackett Laboratory, Department of Physics, Imperial College London, London SW7 2AZ, UK*

*[*j.pendry@imperial.ac.uk](mailto:j.pendry@imperial.ac.uk)*

**Abstract:** Time dependent systems in general do not conserve photons nor do they conserve energy. However when parity-time symmetry holds Maxwell's equations can sometimes both conserve photon number and energy. Here we show that photon conservation is the more widely applicable law which can hold in circumstances where energy conservation is violated shedding further light on an amplification mechanism identified in previous papers as a process of conserved photons climbing a frequency ladder.

## 1. Introduction

Metamaterials are a long established feature in the electromagnetic landscape: fine scale structure in a material can radically alter its response to radiation. More recently interest has been growing in systems structured in time and space [1-6], Usually this is taken to mean that the material itself does not move but that its local properties are modulated in time. Even the simplest space-time structures such as gratings have shown remarkable properties quite distinct from their static counterparts [7-9]. Here we take one further step by asking if there are quantum mechanical consequences of space-time modulation, specifically asking whether the number of photons is conserved. Previous papers [10-12] proved conservation laws by invoking conservation of lines of force which holds under rather specific conditions. In this paper we give a more general proof derived directly from Maxwell's equations.

Our model is a system in which all six components of the permittivity and permeability tensors are functions of $(gx - \Omega t)$ forming a grating with a spatial period of $2\pi/g$ and a time period of $2\pi/\Omega$. Although the material of the grating is stationary, its profile moves in the $x$ direction with a velocity of $c_g = \Omega/g$.

In systems that are non-Hermitian real eigenvalues may nevertheless be recovered if there is symmetry under inversion of both parity and time, formally known as 'PT symmetry'. For example in a PT symmetric system this would imply the existence of eigenvectors with real wave vectors and eigenvalues at real frequencies carrying the implication that for an individual eigenvector energy is conserved over one grating period in space or time. Conservation of the number of photons is an obvious corollary. We shall show that the latter result is more profound and holds under circumstances where energy conservation does not.

It is little remarked that for a *linear combination* of eigenvectors energy is not conserved over one grating period even if all eigenvalues are real, and in some pathological instances [13] energy content may grow exponentially in time even though the eigenvalues remain real. The cause of this anomaly is an orthogonality relationship between eigenvectors which implies that each eigenvalue contributes independently to the total number of photons, *but not to the total energy content*. Our results hold provided that we count photons of negative frequency as having negative energy. Further discussion of this point follows in the penultimate section.

## 2. The transfer matrix and its eigenvectors

We examine photon conservation in two specific instances. In the first case the grating exists throughout all space but only for a finite period of time and we prove that the system contains the same number of photons before and after the grating has been turned on and off. In the second case the system is constant in time but occupies a finite region of space and here we show that the flow of photons out equals the flow into the grating. The finite nature of each of these systems is important and avoids instabilities and singularities that can occur in systems infinite in space and time.

We choose to examine *s*-polarised fields,

$$\mathbf{E} = E_z \hat{\mathbf{z}}, \quad \mathbf{H} = H_y \hat{\mathbf{y}} + H_x \hat{\mathbf{x}} \tag{1}$$

and look for solutions of the form,

$$E_z = \sum_n E_n f(x,y,t), \quad H_y = \sum_n H_{yn} f(x,y,t),$$
$$H_x = \sum_n H_{xn} f(x,y,t), \quad f(x,y,t) = e^{i(k+ng)x + ik_y y - i(\omega+n\Omega)t} \tag{2}$$

On substituting into Maxwell's equations we have the following relations,

$$(k+n'g)H_{yn'} - k_y H_{xn'} + \varepsilon_0 (\omega+n'\Omega) \sum_n \varepsilon_{zn'n} E_n = 0,$$
$$(k+n'g)E_{n'} + \mu_0 (\omega+n'\Omega) \sum_n \mu_{yn'n} H_{yn} = 0, \tag{3}$$
$$k_y E_{n'} - \mu_0 (\omega+n'\Omega) \sum_n \mu_{xn'n} H_{xn} = 0$$

We have taken advantage of the relationship,

$$\frac{\Omega}{2\pi} \int_0^{2\pi/\Omega} e^{-i(k+n'g)x+ik_y y-i(\omega+n'\Omega)t} \frac{\partial}{\partial t}\left[\varepsilon e^{i(k+ng)x+ik_y y-i(\omega+n\Omega)t}\right] dt$$
$$= -\frac{\Omega}{2\pi}(\omega+n'\Omega) \int_0^{2\pi/\Omega} e^{+i(\omega+n'\Omega)t} \varepsilon e^{-i(\omega+n'\Omega)t} dt = -\frac{\Omega}{2\pi}(\omega+n'\Omega)\varepsilon_{n'n} \tag{4}$$

to eliminate explicit dependence on time derivatives of the permittivity and permeability.

## 2.1 A spatially infinite grating existing for a finite time

In this case the system is strictly periodic in space and we expand waves in terms of eigenvectors having different frequencies but the same wave vector modulo $g$. Eq. (3) can be written as a generalised eigenvalue equation,

$$\mathbf{0} = \begin{bmatrix} -\mathbf{W}\boldsymbol{\varepsilon}_z & -c_0(k\mathbf{I}+g\mathbf{N}) & +c_0 k_y \mathbf{I} \\ -c_0(k\mathbf{I}+g\mathbf{N}) & -\mathbf{W}\boldsymbol{\mu}_y & 0 \\ c_0 k_y \mathbf{I} & 0 & -\mathbf{W}\boldsymbol{\mu}_x \end{bmatrix} \begin{bmatrix} \mathbf{E}' \\ \mathbf{H}_y \\ \mathbf{H}_x \end{bmatrix} = \mathbf{M} \begin{bmatrix} \mathbf{E}' \\ \mathbf{H}_y \\ \mathbf{H}_x \end{bmatrix} \tag{5}$$

where $\mathbf{I}$ is the identity matrix, $N_{n'n} = n\delta_{n'n}$, $W_{n'n} = \delta_{n'n}(\omega + n'\Omega)$ and $E'_n = Z_0^{-1} E_n$. The left eigenvalue obeys,

$$\mathbf{0} = \mathbf{W} \begin{bmatrix} -\boldsymbol{\varepsilon}_z \mathbf{W} & -c_0(k\mathbf{I}+g\mathbf{N}) & +c_0 k_y \mathbf{I} \\ -c_0(k\mathbf{I}+g\mathbf{N}) & -\boldsymbol{\mu}_y \mathbf{W} & 0 \\ c_0 k_y \mathbf{I} & 0 & -\boldsymbol{\mu}_x \mathbf{W} \end{bmatrix} \mathbf{W}^{-1} \mathbf{W} \begin{bmatrix} \mathbf{E}'^*_L \\ \mathbf{H}^*_{yL} \\ \mathbf{H}^*_{xL} \end{bmatrix}$$

$$= \begin{bmatrix} -\mathbf{W}\boldsymbol{\varepsilon}_z & -c_0(k\mathbf{I}+g\mathbf{N}) & +c_0 k_y \mathbf{I} \\ -c_0(k\mathbf{I}+g\mathbf{N}) & -\mathbf{W}\boldsymbol{\mu}_y & 0 \\ c_0 k_y \mathbf{I} & 0 & -\mathbf{W}\boldsymbol{\mu}_x \end{bmatrix} \mathbf{W} \begin{bmatrix} \mathbf{E}'^*_L \\ \mathbf{H}^*_{yL} \\ \mathbf{H}^*_{xL} \end{bmatrix} = \mathbf{M} \begin{bmatrix} \mathbf{E}' \\ \mathbf{H}_y \\ \mathbf{H}_x \end{bmatrix} \tag{6}$$

and hence,

$$\begin{bmatrix} \mathbf{E}'^*_L \\ \mathbf{H}^*_{yL} \\ \mathbf{H}^*_{xL} \end{bmatrix} = \mathbf{W}^{-1} \begin{bmatrix} \mathbf{E}' \\ \mathbf{H}_y \\ \mathbf{H}_x \end{bmatrix} \tag{7}$$

It follows that for two different frequencies, $\omega_l, \omega_m$,

$$\mathbf{0} = \begin{bmatrix} \mathbf{E}'^*_m & \mathbf{H}^*_{ym} & \mathbf{H}^*_{xm} \end{bmatrix} \mathbf{W}^{-1} (\mathbf{M}_l - \mathbf{M}_m) \begin{bmatrix} \mathbf{E}'_l \\ \mathbf{H}_{yl} \\ \mathbf{H}_{xl} \end{bmatrix}$$

$$= (\omega_l - \omega_m) \begin{bmatrix} \mathbf{E}'^*_m & \mathbf{H}^*_{ym} & \mathbf{H}^*_{xm} \end{bmatrix} \mathbf{W}^{-1} \begin{bmatrix} -\boldsymbol{\varepsilon}_z & 0 & 0 \\ 0 & -\boldsymbol{\mu}_y & 0 \\ 0 & 0 & -\boldsymbol{\mu}_x \end{bmatrix} \begin{bmatrix} \mathbf{E}'_l \\ \mathbf{H}_{yl} \\ \mathbf{H}_{xl} \end{bmatrix} \tag{8}$$

$$= (\omega_l - \omega_m) \begin{bmatrix} \mathbf{E}'^*_m & \mathbf{H}^*_{ym} & \mathbf{H}^*_{xm} \end{bmatrix} \mathbf{W}^{-1} \begin{bmatrix} \varepsilon_0^{-1} \mathbf{D}'_l \\ \mu_0^{-1} \mathbf{B}_{yl} \\ \mu_0^{-1} \mathbf{B}_{xl} \end{bmatrix}$$

which leads to the following orthogonality relationship,

$$\sum_n \frac{E^*_{mn} D_{ln} + H^*_{ymn} B_{yln} + H^*_{xmn} B_{xln}}{2\hbar(\omega_m + n\Omega)} = \delta_{lm} \tag{9}$$

Where the numerator is twice the energy in each component of the field and the denominator normalises to the number of photons in each eigenvector, assuming for the moment that photons of negative frequency have negative energy.

Photon content of the eigenvector does not change with time if $\omega_m$ is real. Furthermore any linear combination of eigenvectors each with real $\omega_m$ also conserves photon number: in consequence of the orthogonality theorem each Bloch wave contributes independently to the number of photons. As remarked earlier, there is no such conservation theorem for energy in a combination of eigenvectors.

*2.2 A spatially finite grating existing for a infinite time*

In contrast to the first case this grating is strictly periodic in time and therefore we expand waves in terms of eigenvectors having different wave vectors but the same frequency modulo $\Omega$. Two of the eigenvalues of (5) will always be trivial longitudinal modes with $\omega = 0$ and can be eliminated along with $H_x$ to yield a second matrix equation having the wave vector as an eigenvalue,

$$k \begin{bmatrix} \mathbf{E}' \\ \mathbf{H}_y \end{bmatrix} = - \begin{bmatrix} g\mathbf{N} & c_0^{-1} \mathbf{W} \boldsymbol{\mu}_y \\ c_0^{-1} \mathbf{W} \boldsymbol{\varepsilon}_z - k_y^2 c_0 \boldsymbol{\mu}_x^{-1} \mathbf{W}^{-1} & g\mathbf{N} \end{bmatrix} \begin{bmatrix} \mathbf{E}' \\ \mathbf{H}_y \end{bmatrix} \tag{10}$$

This matrix has left eigenvectors obeying,

$$k \begin{bmatrix} \mathbf{E}'_L & \mathbf{H}_{yL} \end{bmatrix} = - \begin{bmatrix} \mathbf{E}'_L & \mathbf{H}_{yL} \end{bmatrix} \begin{bmatrix} g\mathbf{N} & c_0^{-1} \mathbf{W} \boldsymbol{\mu}_y \\ c_0^{-1} \mathbf{W} \boldsymbol{\varepsilon}_z - k_y^2 c_0 \boldsymbol{\mu}_x^{-1} \mathbf{W}^{-1} & g\mathbf{N} \end{bmatrix} \tag{11}$$

an equation that can be manipulated to give,

$$k \begin{bmatrix} \mathbf{W} \mathbf{H}_{yL} & \mathbf{W} \mathbf{E}'_L \end{bmatrix} = - \begin{bmatrix} \mathbf{W} \mathbf{H}_{yL} & \mathbf{W} \mathbf{E}'_L \end{bmatrix} \begin{bmatrix} g\mathbf{N} & c_0^{-1} \boldsymbol{\varepsilon}_z \mathbf{W} - k_y^2 c_0 \mathbf{W}^{-1} \boldsymbol{\mu}_x^{-1} \\ c_0^{-1} \boldsymbol{\mu}_y \mathbf{W} & g\mathbf{N} \end{bmatrix} \tag{12}$$

Finally we take the transpose,

$$k^* \begin{bmatrix} \mathbf{W} {H^*}_{yL} \\ \mathbf{W} \mathbf{E}^*_L \end{bmatrix} = - \begin{bmatrix} g\mathbf{N} & c_0^{-1} \mathbf{W} \boldsymbol{\mu}_y \\ c_0^{-1} \mathbf{W} \boldsymbol{\varepsilon}_z - k_y^2 c_0 \boldsymbol{\mu}_x^{-1} \mathbf{W}^{-1} & g\mathbf{N} \end{bmatrix} \begin{bmatrix} W H^*_{yL} \\ W E^*_L \end{bmatrix} \tag{13}$$

which, provided that $k = k^*$, is the original eigenvector equation. We conclude,

$$\begin{bmatrix} \mathbf{H}_{yL} \\ \mathbf{E'}_L \end{bmatrix} = \begin{bmatrix} \mathbf{W}^{-1}\mathbf{E'}^* \\ \mathbf{W}^{-1}\mathbf{H}_y^* \end{bmatrix} \qquad (14)$$

and we have an orthonormal set.:

$$\delta_{lm} = \frac{Z_0}{2\hbar} \sum_n \frac{E'_{nl} H^*_{ynm} + H_{ynl} E'^*_{nm}}{(\omega + n\Omega)} = \sum_n \frac{E_{nl} H^*_{ynm} + H_{ynl} E^*_{nm}}{2\hbar(\omega + n\Omega)} \qquad (15)$$

where the subscripts $l, m$ refer to eigenvectors with eigenvalues $k_l, k_m$. This normalisation can be identified with a flow of one photon per second: the numerator on the rhs of (15) contains the Poynting vector and corresponds to twice the energy flow, the denominator to twice the energy of each photon. In consequence of (15) each eigenvector contributes independently to the flow of photons and therefore any linear combination of eigenvectors conserves photons. Once more we emphasis that this is not the case for energy, the amplifying nature of this system being well documented: see [13] and references therein.

## 3. Discussion

### 3.1 Guaranteed photon conservation

Our theorems require that the eigenvalues be real: any imaginary values of frequency or wave vector will add or remove photons. However, as was shown in a previous paper, when waves travel in the same direction as the grating, $k_y = 0$, real eigenvalues can be guaranteed by eliminating back scattering and thus closing all band gaps. This is achieved by impedance matching the system so that permittivity and permeability are modulated in tandem and impedance is constant everywhere. At first this would appear to present a paradox because exponential growth of the energy content can still be observed in the so called 'transluminal region' where the grating speed is comparable to the velocity of light. The paradox is resolved by observing that energy is added by requiring the fixed number of photons to climb a ladder of frequencies.

Numerical simulations are presented below: we consider a grating with unit flux of photons incident defined by the following form of the permittivity and permeability.

$$\begin{aligned} \varepsilon(x - c_g t) &= 1 + 2\alpha_\varepsilon \cos(gx - \Omega t) \\ \mu(x - c_g t) &= 1 + 2\alpha_\mu \cos(gx - \Omega t) \end{aligned} \qquad (16)$$

In case A the conditions for exact conservation are met whereas for the other three cases first impedance matching is broken in case B, then the angle of incidence deviates from the normal by $30°$. Other parameters are given in the caption to fig. 1. Note that the conservation law is insensitive to these violations. Table 1 shows how well conservation holds up even with quite extreme deviation from the ideal parameters. Fig. 1 plots the energy flux for each example.

Note the large compression of emerging pulses of energy showing that photons are being distributed over a very wide range of Fourier components. For the parameters chosen our calculations are in the transluminal region where exponential growth of net energy flux occurs despite no more photons being added to the system

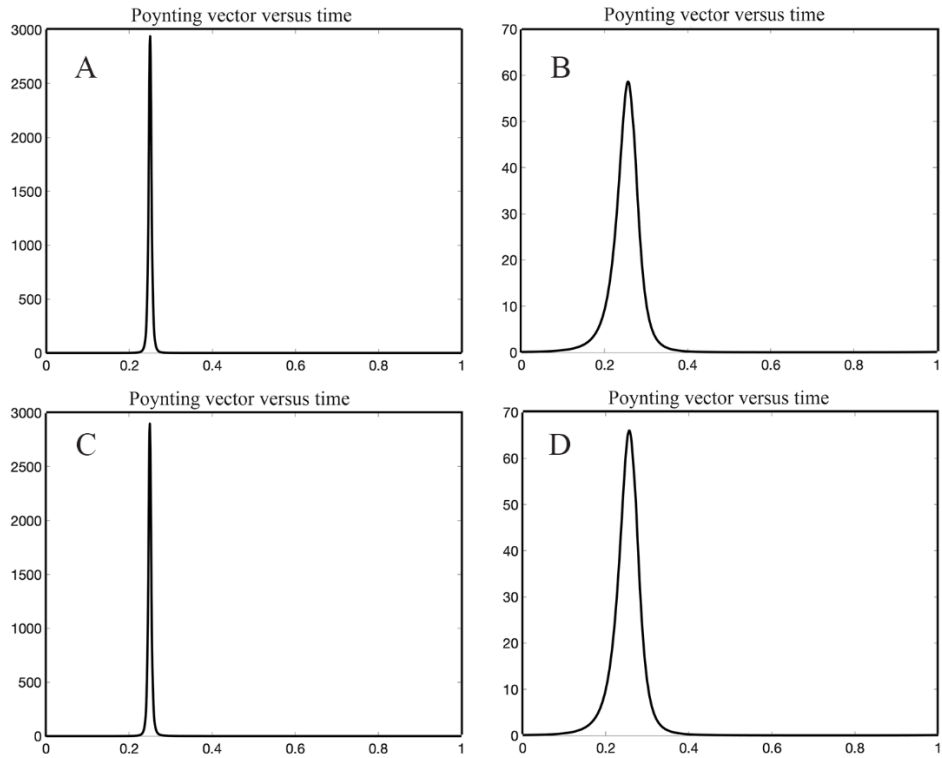

Fig. 1. Flow of energy at the exit to a grating as a function of time normalised to unit incident flux. The abscissa covers one time period of the grating. In each case $\Omega = g = 1.0$, $\omega = 0.3$, $d = 10$, otherwise:

A: $k_y = 0.0$, $\alpha_\varepsilon = 0.2$, $\alpha_\mu = 0.2$
B: $k_y = 0.0$, $\alpha_\varepsilon = 0.2$, $\alpha_\mu = 0.0$
C: $k_y = 0.15$, $\alpha_\varepsilon = 0.2$, $\alpha_\mu = 0.2$
D: $k_y = 0.15$, $\alpha_\varepsilon = 0.2$, $\alpha_\mu = 0.0$

**Table 1.** Numerical simulation of conservation of photon number for various parameters. Conditions for strict conservation are met in case A but violated to varying degrees in the three other examples. A wave-vector of $k_y = 0.15$ corresponds to an angle of incidence $30°$ away from the normal. The labels A, B, C, D, correspond to the labels in fig. 1 where the energy flux is shown as a function of time.

| | |
|---|---|
| A: $k_y = 0.0$, $\alpha_\varepsilon = 0.2$, $\alpha_\mu = 0.2$ | $\tilde{N}_{out}/\tilde{N}_{in} = 1.00000$ |
| B: $k_y = 0.0$, $\alpha_\varepsilon = 0.2$, $\alpha_\mu = 0.0$ | $\tilde{N}_{out}/\tilde{N}_{in} = 1.00034$ |
| C: $k_y = 0.15$, $\alpha_\varepsilon = 0.2$, $\alpha_\mu = 0.2$ | $\tilde{N}_{out}/\tilde{N}_{in} = 0.99946$ |
| D: $k_y = 0.15$, $\alpha_\varepsilon = 0.2$, $\alpha_\mu = 0.0$ | $\tilde{N}_{out}/\tilde{N}_{in} = 0.99984$ |

*3.2 Negative frequency negative energy?*

No, of course not: the energy of a photon is always given by $\hbar|\omega|$, so that if we wish to count real photons rather than imagined entities with negative energies we must use,

$$\sum_{nk} \frac{E_{nk} H^*_{ynk} + H_{ynk} E^*_{nk}}{2\hbar|\omega + n\Omega|} \tag{17}$$

which provokes the question of 'why do our theorems work in terms of putative negative energies?'. Provided that the frequencies concerned are all positive or all negative the question does not arise. If the grating excites photons across the $\pm$ divide then we have to conclude that photons are not conserved because in the sum rule some of the photons enter with the wrong sign, the implication being that photons are not conserved. In fact non-conservation happens not singly but in pairs of photons: to make the sum rule work we must take away a qubit comprising two photons, one with negative frequency, the other with positive frequency,

$$|\hbar\omega| - (|\hbar\omega| + |\hbar\omega|) = -|\hbar\omega| \tag{18}$$

which has implications for spontaneous emission from the vacuum to be discussed in a subsequent paper.

Dirac confronted a similar conundrum posed by his relativistic equations for the electron which show states at negative energy mirroring those at positive energies. His elegant solution was to postulate that in the vacuum state all negative energy states are occupied and excitations consist of electron-positron pairs, in the process identifying a new particle as a 'hole' in the negative energy states. Unfortunately this solution is not open to us as we are dealing with Bosons not Fermions. Nevertheless it does appear that negative and positive frequency photon states are in some way distinct entities. It is often said that the photon is its own antiparticle but this appears to be the case only if both photons have differing signs to their frequencies.

## 4.  Conclusions

Energy conservation in Hermitian systems also implies conservation of number of photons. Here we have shown that time dependence decouples energy conservation form photon conservation. Even in PT symmetric systems where individual eigenstates have real frequencies and conserve both energy and photons, linear combinations conserve only photons and violate energy conservation.

**Funding.** from the Gordon and Betty Moore Foundation.

**References**
1. Christophe Caloz, and Zoé-Lise Deck-Léger, "Spacetime Metamaterials", IEEE Transactions on Antennas and Propagation, **68,** 1569-1582 *and* 1583-1598 (2020).
2. E. Galiffi et al., "Photonics of Time-Varying Media" Advanced Photonics **4**, 014002-01-32 (2022).
3. Joshua N. Winn, Shanhui Fan, John D. Joannopoulos, and Erich P. Ippen, "Interband transitions in photonic crystals", Phys. Rev. B **59**, 1551 (1999).
4. Zongfu Yu, and Shanhui Fan, "Complete optical isolation created by indirect interband photonic transitions", Nature photonics **3,** 91-94 (2009).
5. Dimitrios L. Sounas and Andrea Alù, "Non-reciprocal photonics based on time modulation", Nature Photonics **11,** 774–783 (2017).
6. D. Ramaccia, A. Toscano, F. Bilotti, D. L. Sounas and A. Alù, "Space-time modulated cloaks for breaking reciprocity of antenna radiation", *2019 IEEE International Symposium on Antennas and Propagation and USNC-URSI Radio Science Meeting*, Atlanta, GA, USA, 1607-1608 (2019)
7. E. Galiffi , P. A. Huidobro, and J. B. Pendry, "Broadband Nonreciprocal Amplification in Luminal Metamaterials", Physical Review Letters, **123**, 206101 (2019).
8. Zoé-Lise Deck-Léger,, Nima Chamanara, Maksim Skorobogatiy, Mário G. Silveirinha and Christophe Caloz "Uniform-velocity spacetime crystals" Advanced Photonics **1**, 056002 (2019).
9. Yonatan Sharabi, Alex Dikopoltsev, Eran Lustig, Yaakov Lumer, Mordechai Segev, "Spatiotemporal photonic crystals", Optica, **9**, 585-592 (2022).
10. J.B. Pendry, E. Galiffi, and P. A. Huidobro, "Gain in time-dependent media - a new mechanism", JOSA B, **8**, 3360-3366 (2021).
11. J.B. Pendry, E. Galiffi, and P. A. Huidobro, "Gain mechanism in time-dependent media", Optica, **8**, 636-637 (2021).
12. J.B. Pendry, E. Galiffi, and P. A. Huidobro, "Photon conservation in trans-luminal metamaterials", Optica, **9**, (2022)
13. E Galiffi, MG Silveirinha, PA Huidobro, JB Pendry, "Photon localisation and Bloch symmetry breaking in luminal gratings', *Phys. Rev. A* **104**, 013509 (2021).